
\input harvmac

\def\square{\kern1pt\vbox{\hrule height 1.2pt\hbox{\vrule width 1.2pt\hskip 3pt
   \vbox{\vskip 6pt}\hskip 3pt\vrule width 0.6pt}\hrule height 0.6pt}\kern1pt}
\def\newpage{\vfill\eject}

\def\fun#1#2{\lower3.6pt\vbox{\baselineskip0pt\lineskip.9pt
  \ialign{$\mathsurround=0pt#1\hfil##\hfil$\crcr#2\crcr\sim\crcr}}}

\def\om{\omega}
\def\t{\Theta}

\def\p{\Phi}
\def\P{\Phi}
\def\phi{\Phi}

\def\part{\partial}
\def\ep{\epsilon}

\def\lta{\mathrel{\spose{\lower 3pt\hbox{$\mathchar"218$}}
     \raise 2.0pt\hbox{$\mathchar"13C$}}}
\def\gta{\mathrel{\spose{\lower 3pt\hbox{$\mathchar"218$}}
     \raise 2.0pt\hbox{$\mathchar"13E$}}}
\def\spose#1{\hbox to 0pt{#1\hss}}

\hoffset=1truein
\voffset=1truein
\vsize=23.0truecm
\hsize=16.25truecm
\parskip=0.25truecm
\def\newpage{\vfill\eject}

\def\fun#1#2{\lower3.6pt\vbox{\baselineskip0pt\lineskip.9pt
  \ialign{$\mathsurround=0pt#1\hfil##\hfil$\crcr#2\crcr\sim\crcr}}}
\baselineskip=24truept
\def\P{\Phi}
%
%
$\,$
\vskip 0.75truein
\centerline{\bf  A POSSIBLE SOLUTION TO THE HORIZON
PROBLEM:}
\centerline{\bf THE MAD ERA FOR MASSLESS SCALAR THEORIES OF
GRAVITY}
\vskip 0.2truein
\vskip 0.1truein
\centerline{ {\bf Janna  J. Levin}}
\centerline{\it Department of Physics, MIT}
\centerline{\it Cambridge, MA 02139}
\vskip 0.2truein
\centerline{ {\bf Katherine Freese}$^1$ }
\centerline{\it Physics Department, University of Michigan}
\centerline{\it Ann Arbor, MI 48109}
\centerline{{\rm and} {\it Institute for Theoretical Physics}}
\centerline{\it Santa Barbara, CA 93106}

\vskip 0.4truein

\centerline{\bf ABSTRACT}
\vskip 0.2truein

Extensions of Einstein gravity which allow the gravitational
constant $G$ to change with time as the universe evolves
may provide a resolution to the horizon
problem without invoking
a period of vacuum domination and without the subsequent
entropy violation.
In a cosmology for which the gravitational constant
is not in fact constant, the universe
may be older at a given temperature than in
a standard Hot Big Bang universe;
thus, larger
regions of space could have come into causal
contact at that temperature.
This opens the possibility that large regions became
smooth at some high temperature without violating causality.
The extra aging of the universe can be accomplished by
an early period with a large Planck mass,
a period we call the MAD  era (Modified Aging era or the
Massively Aged and
Detained era). We discuss in this paper theories of gravity
in which the gravitational constant is replaced with
a function of a scalar field.
However, this resolution to the smoothness problem
can more generally be a feature of any physics which allows
the Planck mass to vary with time.
In this paper, we examine scalar theories
of gravity without a potential for the scalar field.
We first consider the original Brans-Dicke proposal and then
address more general scalar theories.

Solutions to the equations of motion for Brans-Dicke
gravity during the radiation dominated era
are presented.  In particular, we study
the evolution of the Brans-Dicke field $\p$
which determines the Planck mass at any given time,
$\p (t) = m_{pl}(t)^2$.  We find that,
regardless of initial conditions for the
Planck mass, it evolves towards an
asymptotic value $\tilde m_{pl} = \tilde \p^{1/2}$.
The same asymptotic behavior is found in
more general scalar theories.
For both a Brans-Dicke cosmology and a more general scalar theory,
the smoothness of the universe can be explained
if the Planck mass is large at some high temperature $T_c$
prior to matter-radiation equality:
specifically, if $\tilde m_{pl}/M_o\gta T_c/T_o$,
where $M_o = 10^{19}$ GeV is the Planck mass today and $T_o$ is
the temperature of the cosmic background radiation today.
In a pure Brans-Dicke cosmology,
an additional mechanism
(e.g. a potential for $\p$) is required
to drive the Planck mass
to the value $M_o$ by today.
In a more general scalar theory of gravity
with variable Brans-Dicke parameter, the suggestion
is made that
the Planck mass may approach the value $M_o$ more rapidly
during the matter dominated era than in a Brans-Dicke cosmology.

\vskip 0.2truein

\centerline{submitted to {\it Physical Review D}}
\centerline{\it November 6, 1992}
\vskip 0.20truein
\centerline{$^1$Alfred P. Sloan Foundation Fellow}
\centerline{Presidential Young Investigator}

\newpage

\vskip 20truept

\centerline{\bf I) Introduction}

The standard Hot Big Bang model of the early universe
is unable to explain the smoothness or flatness of the
observed universe.  In the standard cosmology,
our present horizon volume
would envelop many regions which were causally disconnected
at earlier times.  Consequently, the
homogeneity and isotropy of the observed universe
is a mystery.  Regions which could not have
been in causal contact at earlier times seem nonetheless
to be identical in temperature and other properties, as
the isotropy of the
cosmic background radiation attests.

The inflationary model proposed by Guth$^1$
addresses the horizon and flatness problems, as well
as the monopole problem (if it exists).
As a general class of early universe
models, inflation suggests that
our universe passes
through an era of
false vacuum domination during which the scale factor grows
exponentially
(or at least superluminally).
The superluminal growth of the scale factor
inflates a region which was initially  subhorizon-sized and therefore
in
causal contact.
If the scale factor
grows sufficiently, our observable universe fits inside one of these
blown up causally connected volumes.
During inflation,
the temperature of the universe plunges,
$T\propto R^{-1}$, where $R$ is the scale
factor.  Therefore, the next crucial ingredient for a
successful inflationary
model is a period of entropy violation which reheats the universe to
some high temperature.

In this paper, we propose that a cosmology with a variable Planck mass
can resolve the horizon problem without
a period of vacuum domination.  Further, entropy  is always
conserved. [In another paper, we illustrate
how our model can resolve the monopole
problem
as well; the flatness
problem may be alleviated by our model but is
quite complicated and will be studied elsewhere.]
We have considered (a) (in Section III)
the Brans-Dicke proposal to replace the constant
Planck mass with a scalar field, $m_{pl}\propto \psi$, and
(b) (in Section IV) more general scalar theories
where the Planck mass could be an arbitrary function
of a scalar field $\psi$.
In both cases,
the energy density of the universe begins radiation dominated and
then goes over to a period of matter domination as in the standard
cosmology.

We derive below the analytic
solutions to the cosmological equations of motion
for these alternate theories of gravity when the
energy density in ordinary matter is radiation dominated.
We find the scale factor, the temperature,
the Hubble constant, and the horizon radius in terms of
the variable Planck
mass.  We also find the time evolution
of $m_{pl} (t)$ for early and late times
during the radiation dominated era.

The initial time derivative of the Planck mass,
$d m_{pl}/dt$, may be positive, negative, or zero.
Though the specifics of the cosmology depend on the initial conditions,
there is a common nature to the solutions.
If $d m_{pl}/dt<0$ initially, then the Planck mass starts large and
quickly drops to some asymptotic value, denoted $\tilde m_{pl}$.
If $d m_{pl}/dt>0$ initially, then the Planck mass starts out
smaller than $\tilde m_{pl}$
and again quickly approaches the asymptotic value $\tilde m_{pl}$.
In either case, as long as
$d m_{pl}/dt$ is significant, the scale factor and the temperature
evolve with the changing $m_{pl}$ in a complicated way.
Thus, even though the energy density in ordinary matter is
radiation dominated,
the variation in the Planck mass alters the dynamics from that
of a standard radiation dominated cosmology.
However, once $m_{pl}$ veers close to its asymptotic value $\tilde m_{pl}$,
then $d m_{pl}/dt\approx 0$ and the universe evolves in a familiar way.
At this point, the descriptions of the cosmology for the
three possible initial $dm_{pl}/dt$ (positive, negative, or zero)
converge.  The equations of motion reduce to those of an ordinary
radiation dominated Einstein cosmology
with $M_o$, the usual Planck mass of $10^{19}$ GeV, replaced with $\tilde
m_{pl}$.
In particular, this means $R\propto \tilde m_{pl}^{-1/2} t^{1/2}$,
$H=1/2t$, and $d_{\rm horiz}\propto t$
where $R$ is the scale factor, $H$ is the Hubble constant,
and $d_{\rm horiz}$ is the horizon radius.
Thus, despite the underlying structure of the theory,
gravity appears to be described by general relativity
with a static gravitational constant.

However, in this early phase of the universe,
the gravitational constant can be much larger
than its value today:  $\tilde m_{pl} \gg M_o$.
We call this epoch of large Planck mass
the MAD epoch.
Here we illustrate briefly how a MAD
expansion can resolve the horizon problem.
Once the Planck mass has reached its asymptotic value $\tilde
m_{pl}$,
the age of the
universe scales as $t \propto \tilde m_{pl}/T^2$.
A universe which has
an early  MAD  era with
large Planck mass
$\tilde m_{pl}>>M_o$ is therefore older at a given temperature
than a universe which has today's value of the Planck mass
$M_o$ for all time.
This gives us a hint as to how such a scenario may solve the smoothness
problem.
If the universe is older than in the standard model,
then much larger regions of spacetime would
have come into contact than we had previously supposed.
This opens the possibility that large regions
became smooth without violating causality.
We will describe this approach to resolving the
horizon problem
more quantitatively later.

Before we move on to the calculation, we should complete the
history of this cosmology.  As in a standard cosmology, the
energy density in nonrelativistic matter will eventually
exceed the energy density in radiation.  Thus the era of radiation domination
will end as the universe becomes matter dominated.
A matter dominated Brans-Dicke cosmology has been well studied$^2$,
as have the constraints on such models$^3$.
We discuss the constraints on MAD  expansion
with Brans-Dicke gravity in section III.D.
We mention here one of the most severe constraints
on our model.  As discussed above,
in order to solve the smoothness problem,
our model requires a large value of the Planck
mass $\tilde m_{pl}$ at some time during
the radiation dominated epoch.
During the matter dominated era, the Planck mass
will continue to evolve.  However, if the Brans-Dicke parameter
is greater than 500 as the observations imply, then
in pure Brans-Dicke gravity,
$m_{pl}$ will not evolve enough during matter domination
to reach the value of $M_o$
today.  We suggest additional physics which may amend this
problem, such as a potential for the $\psi$
field or a more general scalar theory of gravity.

We consider a more general class of scalar theories
in which the Brans-Dicke parameter $\omega$ is not
constant, still without a potential for
the scalar field.
We present solutions during the radiation dominated era.
In this case of variable $\omega$, the Planck mass
can change more rapidly during the matter-dominated
era and could conceivably
reach the value $M_o$ by today.
To check this conjecture, the solutions to the equations of
motion during a matter dominated era for a  general scalar
theory would have to be obtained.  In addition, the constraints
on more general models need to be studied
in future work.

Extended inflation$^4$ and hyperextended inflation$^5$ were both developed in
the context of scalar theories of gravity.  In addition to the scalar
field which couples to gravity, these models require another scalar field,
the inflaton field,
and a potential for the inflaton.
The horizon problem is resolved in the usual inflationary way as the
scale factor grows superluminally during an era of false vacuum
domination and then the universe is reheated during a period of
entropy violation.
It is interesting to note that these models also
need additional mechanism,
such as a potential for the Brans-Dicke field,
to drive the Planck mass down to the value $M_o$ by today.

In Section II) we present the action and equations
of motion for the alternate theories of gravity
that we are considering.
Section III) focuses on Brans-Dicke gravity:
III.A) presents solutions to the equations of
motion during the radiation dominated era,
with solutions parametrized in terms of the
Brans-Dicke field $\p$;
III.B) relates these solutions to time evolution;
III.C) illustrates the causality condition
required to solve the horizon problem;
and III.D) discusses problems with and
constraints on the scenario.
Section IV) presents a general
(albeit somewhat preliminary) discussion
of the MAD era  in the context of
more general alternate theories
of gravity in which the Brans-Dicke parameter
$\omega$ is not constant.
We summarize our conclusions in Section V).

\vskip .125in
\centerline{\bf II) Action}

Brans and Dicke proposed an extension of Einstein gravity
in which a scalar field
usurps the role assumed by
the gravitational constant in the Einstein action;
that is, in Brans-Dicke gravity, the gravitational constant $G$ is
not a fundamental constant but is instead inversely
proportional to a scalar field.
More generally, $G$
may be some more
complicated function of a scalar field $\psi$:
$G^{-1} = m_{pl}^2=\Phi(\psi)$.
The most general scalar-tensor theories$^3$
were originally studied by Bergmann$^{6}$
and by Wagoner$^{7}$.
Regardless of the specific form of $\Phi$, the action for such an
extension of general relativity is
  \eqn\je{A=\int d^4x\sqrt{-g}\left[-{\Phi(\psi)\over 16\pi }{\cal R}
        - {\omega \over \Phi}{\partial_{\mu}\Phi
        \partial^{\mu}\Phi
        \over 16\pi} - V(\p(\psi))+{\cal L}_{\rm matter}\right ]\ \ ,
        }
where we have used the metric convention
$(-,+,+,+)$, ${\cal R}$ is the scalar curvature,
${\cal L}_{\rm matter}$
is the Lagrangian density for all the matter fields
excluding the field $\psi$, and $V(\p(\psi))$ is the potential for the
field $\psi$.
The parameter $\omega$ is defined by
$\omega=8\pi{\Phi \over  (\partial \Phi/\partial \psi)^2}$.

Stationarizing this action
in a Robertson-Walker metric
gives the equations of motion
for the scale factor of the universe $R(t)$
and for $\Phi(t)$,
  \eqn\one{\ddot \Phi +3H\dot \Phi={8\pi\over (3+2\omega)}(\rho-3p)
        -{\partial U_{\rm }\over \partial \Phi}
        -{1 \over (3+2\omega)}
        {d\omega\over d\Phi} \dot \Phi^2 }
  \eqn\two{H^2+{\kappa \over R^2}={8\pi(\rho+V(\psi)) \over 3\Phi}
        -{\dot\Phi\over\Phi} H +{\omega \over 6}\left({\dot \Phi
        \over \Phi}\right)^2 }
where
  \eqn\three{
        {\partial U_{\rm  } \over \partial \Phi}={16\pi\over (3+2\omega)}
        \left [\Phi{\partial V \over \partial \Phi}-2V\right ] ; }
$U_{\rm  }$ effectively acts as a potential term in the
equation of motion for $\Phi$.
$H=\dot R/R$ is the Hubble constant, while $\rho$ is
the energy density and
$p$ is the
pressure in all fields excluding the
$\psi$ field.

The energy-momentum tensor of matter, $T^{\mu \nu}_{\rm matter}$,
is conserved independently of
the energy-momentum tensor for the scalar, $T^{\mu \nu}_{\p}$.
The conservation equations are
  \eqn\tu{T^{\mu \nu}_{\rm matter \ ;\nu}=0 }
and
\eqn\il{-8\pi T^{\mu \nu}_{\Phi \ ;\mu}=({\cal R}^{\mu \nu}
        -{1\over 2}g^{\mu\nu}{\cal R})\Phi_{;\mu}\ \ . }
Equation \il \ returns the equation of motion of \one.
It can be shown that in an isotropic and homogeneous universe,
the $\mu=0$ component of equation \tu \ gives
  \eqn\six{\dot \rho =-(\rho+p)3H\ \ \ . }
Consider the radiation dominated era
where $\rho=(\pi^2/30)g_*T^4$, $p=\rho/3$, and
$g_*(t)$ is the number of
relativistic degrees of freedom
in equilibrium
at time $t$.
Since conservation of energy-momentum in ordinary matter
does not involve $\Phi$,
we can deduce from eqn \six \ that the entropy per comoving volume
in ordinary matter, $S=(\rho +p)V/ T$,
is conserved.  For convenience we define
  \eqn\seven{\bar S=R^3T^3 , }
where $S = \bar S (4/3)(\pi^2/30)g_S$ and $g_S$ is the number of
relativistic degrees of freedom contributing to the entropy.
For practical purposes we can take $g_S=g_*$.

Once the equation of state, $p(\rho)$,
and the forms of $\Phi(\psi)$ and $V(\psi)$ are specified, these
equations
describe the evolution of the scale factor, the energy density,
and the Planck mass.
In specific, the
equations of motion \one \ and \two \ and the conservation equation \six \
determine
$\Phi(t)$, $\rho(t)$, and $R(t)$ up to four constants of integration.
Notice that $\bar S$ is the constant of integration
from integrating
the energy equation \six.
We take the other three constants to be
the initial value of $\Phi$, the constant $C$ defined
in eqn. (10) ($C \propto \dot \p$),
and the constant of integration $\tilde \p$
given in eqn. (15) (the asymptotic value
of $\p$ in the radiation dominated era).
Given these four initial conditions,
the entire cosmology is specified:
the equations of motion uniquely determine
$R(t)$, $\p(t)$ and $\rho(t)$ for all time.
In contrast,
the standard cosmology with a constant Planck mass
requires only that
the value of the Planck mass and
two boundary values
be specified.
The two values needed could, for example, be the entropy
and the initial value of the scale factor.

To illustrate how the underlying
structure of the Planck mass can alleviate
the horizon problem, we will present two simple possible scenarios here;
that is, we treat separately two different forms of $\p(\psi)$
in the radiation bath of the early universe where $p(\rho)=\rho/3$.
We first treat the original Brans-Dicke proposal of
$\P={2\pi\over \omega}\psi^2$ with $\omega$, defined above, constant.
The second case we consider is general $\P(\psi)$ for which
$\omega$ is not constant.
In both cases
we take $V(\psi)$=0,
although in
another paper we treat the model with a nonzero
potential.
Still, we want to stress that
an explanation of the smoothness of
our observable universe in a cosmology with a variable Planck mass
is more general than the specific models we study.
The crucial ingredients are forms
of $\Phi(\psi)$ and $V(\psi)$ which resolve the horizon
problem without requiring an
epoch of vacuum domination nor of entropy violation.

\centerline{\bf III) Case a: $\om$=constant}

As a first example, take the original Brans-Dicke model where
$\omega$ is constant and
  \eqn\BD{m_{pl}^2=\Phi={2\pi \over \om}\psi^2 \ \ .}
Also, we take $V(\psi)=0$.
In the limit $\omega = \infty$, Brans-Dicke gravity
is the same as Einstein gravity; here we consider
arbitrary $\omega$ and comment on experimental
bounds on $\omega$ below.
We are interested in the behavior of the solutions during the hot
radiation dominated era of the early universe.
We assume the matter energy density is negligible.
Then $p=\rho/3$ and
$\rho-3p$=0.

We first obtain solutions to the equations of motion.
Instead of finding $\Phi(t)$, $R(t)$, and $T(t)$, it is more
tractable to parameterize $R$, $T$, and hence $H$ and the
horizon radius, $d_{\rm horiz}$, by $\Phi$.
We then find approximate solutions for $\Phi$ as a function of
time in different regimes.  Before moving on to
the second case of $\om\ne$ constant, we address the horizon problem
in the context of our solutions.

\centerline{\bf III. A) Solutions to the Equations of Motion}

With $\om$ constant and $V(\psi)=0$,
the $\Phi$ equation of motion
during the radiation dominated era reduces to
$\ddot \Phi +3H\dot \Phi=0$, so that
  \eqn\nine{
        \dot \Phi R^3=-C\ \ ,\ \ {\rm and} \ \ H=-{\ddot \Phi \over 3 \dot
\Phi}
        \ \ ,}
where $C$ is a constant of integration
which can be positive, negative, or zero.
If $C = 0$, then $\dot \p = 0$,
the Planck mass has a constant value
which we may call $\tilde m_{pl}$,
and the universe evolves in the usual
radiation dominated fashion, but with
$G \propto 1/\tilde m_{pl}^2$.
Note that if $C>0$, then $\dot \Phi < 0$,
while if $C<0$, then $\dot \Phi >0$.
Here we take $\kappa = 0$ to illustrate
the behavior of the solutions.  [In the appendix,
we present the solutions for $\kappa = \pm 1$.]
First we solve equation \two \ for $H$:
  \eqn\ten{H=-{\dot \Phi \over 2\Phi}\left
      [1\pm{\sqrt{1+{2\omega \over 3}
        +{4\bar S^{4/3}\gamma\over R^4}
        \left({\Phi \over \dot \p^2}\right )}}\right ]
        \ \ ,}
where $\gamma(t)=(8\pi/3)(\pi^2/30)g_*(t)$.
Note that all three terms inside the square root are
positive quantities.  We choose the sign
in front of the square root in such a way
as to obtain an expanding universe with $H >0$.
Thus, for $C>0$ we take the
$+$ sign in equation \ten, whereas for $C<0$, we take
the $-$ sign in equation \ten.
Throughout the rest of the paper,
the upper sign in equations will refer to the case $C>0$
and the lower sign to the case $C<0$.
Substituting $\dot \p$ from eqn. \nine \ into the square root
in eqn. \ten \ and
using $H = \dot R/R$,
we have
  \eqn\eleven{{dR \over R} = - {d\p \over 2 \p}
        \left [1 \pm \left (1+ {2 \omega \over 3}
        + 4 \bar S^{4/3} \gamma \p C^{-2} R^2 \right )^{1/2}
        \right ] \ \ .}
We define
  \eqn\twelve{\chi(\p) = 4 \bar S^{4/3} \gamma C^{-2}
        \left ({1 \over 1 + 2 \omega/3}\right) \p R^2 \ \ }
and note that $\chi$ is always a real positive quantity.
The integral of eqn. \eleven \ becomes
  \eqn\fourteen{\int_{\chi_i}^\chi
{d\chi'\over \chi'\sqrt{1+\chi'}}=\mp
        \int_{\Phi_i}^\Phi \left (1+{2\omega'\over 3}\right)^{1/2}
        {d\Phi' \over \Phi'}\ \ ,}
where subscript $i$ refers to initial values.
We find the (positive) solution
  \eqn\fifteen{\chi^{-1/2}=
        \sinh\left\{\ep
        \ln  \left [{\Phi \over \tilde \Phi }\right ]
        \right\}\ \ .}
Here, $\tilde \Phi = \Phi_i \exp \left[-(1/\epsilon)
{\rm arcsinh} (\chi_i^{-1/2})\right]$;
i.e. we have absorbed the constants of integration
(which depend on the initial values of $\Phi$ and $\chi$)
into $\tilde \Phi$.
Here, $\ep \equiv {\pm(1+2\omega/3)^{1/2}\over 2}$ (as noted above,
an expanding universe corresponds to the + in $\ep$ if $C>0$,
or the $-$ in $\ep$ if $C<0$).
For convenience, we define
  \eqn\si{\Theta\equiv
        \ep
        \ln \left [
        {\Phi \over \tilde \Phi }\right ]
        \ \ .}
In many of the equations below we express
the field in terms of $\Theta$ rather than $\Phi$.
As we will show below,
$\Theta$ is always positive semidefinite
for any value of $C$:
it ranges from $\Theta = 0$ for $\Phi = \tilde \Phi$
to $\Theta = \infty$ for $\Phi$ far from $\tilde \Phi$.
{}From eqns. \twelve \ , \fifteen \ and \si \ , we
obtain an expression for the scale factor,
  \eqn\sixt{R(\Phi)={C\over \bar S^{2/3}}
        {\ep \over \gamma^{1/2}}
        \tilde \Phi^{-1/2}
        \exp\left[-{\Theta\over2\ep}\right]
        {1\over
        \sinh\Theta} \ \ .}
Note that the product $(C \ep)$ is always positive
semidefinite.

It follows from adiabaticity that
  \eqn\elle{T(\Phi)={\bar S^{1/3} \over R(\Phi)}={\bar S \over C}
        {\gamma^{1/2}\over \ep}
        \tilde \Phi^{1/2}\exp\left[{\Theta\over2\ep}
        \right]
        \sinh\Theta \ \ .}
 The Hubble constant $H(\Phi)$
is obtained from
  \eqn\eighteen{H(\Phi)=
        {1\over R}{dR \over d\Phi}{d\Phi \over dt}=-{C\over R^4}{dR \over
        d\Phi}\ \ , }
where in the last step we used eqn. \nine \ .
We find
  \eqn\noin{H(\Phi)=
        \left ({\bar S\over C}\right )^{2}
        {\gamma^{3/2}\over \ep^3}
        {\tilde \Phi^{1/2}
        \over 2}
        \exp\left[{\Theta\over2\ep}\right]
        \sinh^2\Theta\left\{\sinh\Theta
        +2\ep\cosh\Theta\right\}\ \ .}
 The comoving horizon size is
  \eqn\horiz{{d_{\rm horiz}\over R(\p)}=\int_0^t{dt'\over d\p'}
        {d\p' \over R(\p')} = {-1\over C} \int_{\p(t=0)}^\p
        R(\p')^2 d\p', }
where we have used $\dot \p=-C/R(\p)^3$
in the last equality.
We can integrate this to find
  \eqn\horizphi{{d_{\rm horiz}\over R(\p)}=\left ({C\over \bar S^{4/3}}\right )
        {\ep\over \gamma}\left[{1 \over \tanh \Theta}
        - {1 \over \tanh \Theta(t=0)}\right]\ \ . }
If $\p(t=0)$ starts out far from $\tilde \p$,
i.e. $\Theta(t=0) \gg 1$, eqn. \horizphi \ becomes
  \eqn\horizfit{{d_{\rm horiz}\over R(\p)}=\left ({C\over \bar S^{4/3}}\right )
        {\ep\over \gamma} {\exp\left [{-\t}\right ]\over \sinh\t} \ \ . }

As discussed above,  $R(\p),\ T(\p),\ H(\p), \ {\rm and}\ d_{\rm horiz}(\p)$
are determined only up to the arbitrary constants $\tilde \p, \ C, \ {\rm and}
\ \bar S$.  For instance, eq. \elle \
shows that one can choose the temperature
at a given value of $\p$ by choosing $C/\bar S$ and $\tilde \p$ appropriately.
The fourth and last constant of integration, $\p(t=0)$,
is determined
when $t(\P)$ is found in section III.B.

Even before we determine $\P(t)$, we can
understand the general sketch of the universe's evolution.
We will find that, in all cases, the field $\Phi$
asymptotically
approaches the value $\tilde \p$:
for $C>0$, $\p$ approaches $\tilde \p$ from above;
whereas for $C<0$, $\p$ approaches $\tilde \p$
from below. [For $C=0$, $\p = \tilde \p$
for all time].

Let us first consider the case of $C>0$.
As mentioned previously, this corresponds
to $\dot \Phi < 0$ and $\epsilon >0$.
In order for the scale factor to satisfy $R \geq 0$,
from \sixt \ we can see that we need $\Theta \geq 0$,
i.e., $\Phi \geq \tilde \Phi$.  In addition,
we need the scale factor to grow in time;
again, this requires $\dot \Phi < 0$.
In short, for $C>0$, $\Phi$ starts larger
than $\tilde \Phi$ and
decreases towards the asymptotic value $\tilde \Phi$.

For the case of $C<0$, we have $\dot \Phi >0$
from \nine \ and $\ep < 0$.
To obtain $R \geq 0$, we need $\Theta \geq 0$,
which in this case corresponds to the opposite
limit of $\Phi \leq \tilde \Phi$.
One can show that as $\Phi$ grows towards
its asymptotic value (i.e. $\Theta$ drops),
as long as $| \epsilon| > 1/2$ (i.e. $\omega >0$),
$dR/d\Theta < 0$; the scale factor
grows in time.  In short, for $C<0$, $\Phi$
starts smaller than $\tilde \Phi$ and
grows towards the asymptotic value $\tilde \Phi$.

As we have seen, as
$\Phi$ approaches $\tilde \Phi$,
for $|\epsilon|>1/2$ ($\omega > 0)$,
$R(\p)$ grows and thus
the temperature drops adiabatically.
In addition, one can show
(again for $|\epsilon| >1/2$) that
the comoving horizon size grows, as does
$H^{-1}R^{-1}$.  We'll see below that
the size of a causally connected region
can grow great enough to resolve the horizon problem.
As the universe cools below the temperature of
matter radiation equality ($T_{eq} \approx 5.5 \Omega_M h^2$ eV,
where $\Omega_M$ is the fraction of the critical density
contributed by matter),
the reign of
radiation yields to that of matter and the
nature of the solutions changes.
During matter domination,
the equation of state is
$p(\rho)=0$, and $\rho-3p=\rho$.  This
alters the dynamics considerably. Thus there is a built in off-switch
to end the radiation dominated
behavior of $R(\p),T(\p)$, $H(\p)$, and $d_{\rm horiz}(\p)$.

We will quantify these statements below and find constraints on some of the
constants of integration needed to resolve the horizon problem.
In the next section, we find
approximate descriptions of the behavior of
$\p$ as a function of $t$.  Actually,
the resolution of the smoothness problem
and the evolution of the cosmology can be understood without knowing
$\p$ as a function of $t$.  The universe will pass through the familiar
stages of baryogenesis, nucleosynthesis, matter domination
etc. as the temperature
passes through the relevant energy scale.
To follow the evolution of the universe
all one needs to know is the temperature as a function of $\p$.
One need not know the actual age of the universe.  Still, to
ground the solutions in a slightly more familiar setting we
will indicate below how the universe evolves in time.

\centerline{\bf III.B) The Age of the Universe}

We will determine the time evolution
of the Brans-Dicke field $\p$ in two
different limits:  $\p$ far from $\tilde \p$ ($\Theta \gg 1$),
and $\p \simeq \tilde \p$ ($\Theta \ll 1$).
As we have seen, initially
$\p$ may be large for
$\dot \p <0$ ($C > 0$) or small
for $\dot \p >0$ ($C >0)$.  While $\p$
is far from its asymptotic value $\tilde \p$,
the term $\dot \Phi / \Phi$ contributes significantly
to the equations of motion (see eqn. \two \ ), and
the evolution of the universe is modified relative
to that of an Einstein universe in a complicated way as eqns.
\sixt-\horizfit \
show.
Once $\p\approx \tilde \p$ however,
the universe evolves with time as an ordinary radiation dominated
cosmology with the Planck mass $M_o$ replaced with
$\tilde m_{pl}=\tilde \P^{1/2}$.

To uncover $\Phi(t)$, return to
$\dot \p=-C/R^3$ (c.f. eqn. \nine \ ).
We integrate this equation to find
  \eqn\ittt{\int_{\Phi(t=0)}^\Phi d\Phi' R^3(\Phi')=-Ct \ \ , }
where $R(\p)$ is given in eqn. \sixt \ .
To get a rough feeling for how $\p$ changes with $t$
we find approximate solutions to this integral for two regimes:
(1) $\Phi$ far from the asymptotic value
$\tilde \Phi$ and (2) $\Phi\approx \tilde \Phi$.

\centerline{ (1) $\p$ far from $\tilde \p$}

First we consider the early regime where $\p$
is far from $\tilde \p$, i.e. $\Theta \gg 1$.
The integral on the left hand side of
eqn. \ittt \ is easiest to evaluate if rewritten
in terms of $\Theta$ rather than $\Phi$.
Then we can approximate $\sinh \Theta \simeq e^\Theta/2$
in evaluating the integral.
The lower limit of the $\Theta$ integral
is determined by the boundary condition: $R(t=0) \rightarrow 0$.
For all values of $C$, this initial
value of the scale factor requires
$\Theta(t=0) \rightarrow \infty$;  i.e.,
$\Phi(t=0)\rightarrow \infty$ for $C>0$ while
$\Phi(t=0) \rightarrow 0$ for $C<0$.  Thus, our
fourth and last integration constant is determined.
[Given $C, \bar S,$ and $\tilde \p, R(t=0)$
and $\p(t=0)$ contain the same information via eqn. \sixt \ ].
We can now evaluate eqn. \ittt \ to find
  \eqn\lala{\Phi\approx \tilde \Phi
        \left [\left ({\bar S\over C}\right )^2
        {\gamma^{3/2}\over 8 \ep^3}(1/2 + 3\ep)\right ]^{-{1\over (1/2+3\ep)}}
        \left(\tilde\Phi^{1/2}t\right)^{-{1\over (1/2+3\ep)}}
        \ \ .}

We see from eqns. \lala \ and
\sixt \ that initially $R(t)\propto t^{(1+2\ep)/(1+6\ep)}$.
Remember $\ep=\pm1/2(1+2\om/3)^{1/2}$.  It is interesting to
consider the nature of these solutions for
large deviations from Einstein gravity (i.e. small $\om$).
Take $C<0$ for
$\om\rightarrow 0$, and
thus $\ep\rightarrow -1/2$. In this limit, $\p\rightarrow t$ while
$R\rightarrow{\rm constant}$. Note that this
behavior of the scale factor could also be seen
directly from eq. \sixt \ .
On the other hand,
for $C>0$ with
$\om\rightarrow 0$, $\ep\rightarrow +1/2$ and
$\p\rightarrow t^{-1/2}$ while $R\rightarrow t^{1/2}$.
Again, the behavior $R \propto \p^{-1}$
could be seen directly from
eqns. \sixt \ and \elle.

\centerline{ (2) $\p\approx \tilde \p$}

The previous approximation breaks down for $\Phi\approx \tilde \Phi$.
Again, it is easiest to work with
$\t=\ep\ln(\P/\tilde\p)$.  When $\p/\tilde \p$ is near 1, then $\t \ll 1$
and $R[\t(\p)] \propto \Theta^{-1}$.
The lowest order contribution to the integral yields
  \eqn\thit{\Theta = \left[ {(\epsilon C)^2 \tilde \p ^{-1/2}
        \over 2 \bar S^2 \gamma^{3/2} }{1 \over t} \right]^{1/2}
        \ \ , }
or, equivalently,
  \eqn\Phitime{\Phi=\tilde \Phi\exp\left[\pm
        \left({C^2\over 2
        \bar S^2\gamma^{3/2}
        \tilde \p^{1/2}t}\right)^{1/2}\right]
        \ \ , }
where the plus sign refers to $C>0$ and the minus sign
to $C<0$.
[By assumption, we are working near $\p\approx\tilde \p$ so that
the exponent must be small for
this approximation to be valid.]

Since eqn. \Phitime \ implies that $\Theta(t) \propto t^{-1/2}$
(times a positive constant), we have $R(t) \propto \Theta^{-1}
\propto t^{1/2}$.
So, as $\p$ approaches $\tilde \p$,
the universe evolves as an ordinary radiation
dominated universe with one modification;
the Planck mass $M_o$ is replaced by $\tilde \p^{1/2}$.

In the standard Hot Big Bang model
described by Einstein gravity,
the age of the universe as a function of temperature is
given by $t_{\rm einst}={M_o \over 2\gamma^{1/2}T^2}$.
As $\P$ approaches $\tilde \P$, we can see
from eqns. \Phitime \ and \elle \ that
the age of the universe as a function
of $T(\p)$ mimics this form,
  \eqn\th{t(\p)={\tilde m_{pl}\over 2 \gamma^{1/2}T(\p)^2}\ \  ,}
where $\tilde m_{pl}=\tilde \p^{1/2}$.  Incidentally, this is exactly
the result one obtains for $C=0$ ($\dot \p=0$) and $\p=\tilde \p$.
As discussed in the introduction, the universe is older
for a given temperature
if $\tilde m_{pl}>M_o$
than an Einstein universe with Planck mass $M_o$.
We will show below that if the comoving horizon volume is to
become smooth at high temperatures, then
$\tilde m_{pl}$ must greatly exceed $M_o$.

\centerline{\bf III.C) Horizon Condition and Discussion}

We show here how a Brans-Dicke, radiation dominated
cosmology can resolve the horizon problem
if there is an early MAD epoch  with
$\tilde m_{pl} \gg M_o$.

To explain the smoothness
of our present universe, a
region causally connected at some early time
must grow big enough by today
to encompass our observable universe.
Since we can only see back to the time of decoupling, or perhaps
nucleosynthesis, we
can take the present comoving Hubble radius,
$1/(H_o R_o)$, as a measure of the
comoving radius of the observable universe; here
$H_o$ is the Hubble constant today and $R_o$ is the scale factor
today.  Then
the smoothness
of the observable universe can be explained if a comoving region of radius
at least as large as $1/H_o R_o$ is in causal contact at some time $t_c$
before nucleosynthesis, i.e.,
  \eqn\horstar{{1 \over H_cR_c}>{1\over R_oH_o}\ \ , }
where subscript $c$ denotes values at the time causality is satisfied.

We can express both $H(\p)$ and $H_o$ in terms of the temperature
and the Brans-Dicke field $\p$.
Substitution of expression \elle \ for $T(\p)$ into equation \noin
\ for $H(\p)$ gives
  \eqn\Hsimple{H(\p)=\gamma^{1/2}{T(\p)^2\over \p^{1/2}}{1\over 2\ep}
        \{\sinh{\Theta}+2\ep\cosh\Theta\}\ \ .}
The Hubble constant today can be written as
  \eqn\Hnot{H_o=\alpha_o^{1/2} {T_o^2\over M_o}}
where $T_o=2.6 \times 10^{-13}$ GeV, $M_o$ is the value
of the Planck mass today, and $\alpha_o =
\gamma(t_o) \eta_o = (8\pi/3)(\pi^2/30)g_*(t_o)\eta_o$,
where $\eta_o\sim 10^{4}-10^5$
is the ratio today of the energy density in matter
to that in radiation.
Also, we use adiabaticity, $RT=\bar S^{1/3}\propto (S/g_*)^{1/3}$,
to write the causality
condition as
  \eqn\hee { { {\Phi_c}^{1/2} \over T_c}
        {2\ep\over \sinh\Theta_c+2\ep\cosh\Theta_c}\gta
        \beta {M_o\over T_o}\ \ , }
where $\beta=(\gamma(t_c)/\alpha_o)^{1/2}(g_*(t_c)/g_*(t_o))^{-1/3}$.
To resolve the horizon problem,
this constraint must be satisfied prior to matter/radiation equality.

Although it is possible for the causality condition \hee \
to be satisfied while $\Phi$ is still far from $\tilde \Phi$,
we find that
the solution to the horizon problem that deviates
the least from Einstein gravity
is obtained for
$\Phi \simeq \tilde \p$ in eqn. \hee \ .  In other words,
the lowest possible value of $\Phi_c^{1/2}
\propto m_{pl}(t_c)$ that solves causality is given
by $\p_c \simeq \tilde \p$. [For $C<0$, $\p$
is always less than $\tilde \p$, yet the previous
statement still holds to better than 1 \%; if $\p
\ll \tilde \p$, both $\p$ and $\tilde \p$ are driven to higher values
than if they are equal to each other].
{}From now on, we will examine the causality constraint
for $\Phi$ near its asymptotic value $\tilde \p$.

For $\p \simeq \tilde \Phi$,
$\Theta \simeq 0$,
$\sinh\Theta\simeq 0$, $\cosh\Theta\simeq 1$,
and the causality condition becomes simply
  \eqn\he{ {\tilde m_{pl}\over M_o}\gta \beta
        {T(\tilde \p)\over T_o}\ \ ,}
where $m_{pl}(t_c)\approx \tilde m_{pl}=\tilde \p^{1/2}$.
We can specify the temperature at which we would like to resolve
the causality dilemma.  We are free to
choose the temperature at which $\p=\tilde \p$ since
this is equivalent to making an appropriate choice for the ratio
of the arbitrary constants $\bar S/C$ (see eq. \elle).
[Since $\bar S/C \propto T^3 / \dot \p$,
this amounts to making a choice for $\dot \p (t_c)$;
in principle one should check that this choice is
consistent with measurements of $\dot G /G$ today.
However, since $\dot \p \propto R^{-3}$,
in many cases the time
derivative may be quite small and therefore
unobservable by the present epoch.]

As an example, we consider $T(\Phi_c\approx \tilde \p) \simeq 1$ MeV,
roughly
the temperature of primordial
nucleosynthesis.
Then condition \he \ requires
        \eqn\imp{\tilde m_{pl}\gta 10^{7}M_o
        \ \ . }

We can verify that the smoothness problem is explained by an
old universe.
We showed with our approximate expressions for $m_{pl}$ as a
function of time,
that when $\p\approx \tilde \p$,
the universe evolves as an ordinary radiation dominated
universe with one modification;
$M_o$ is replaced by $\tilde m_{pl}$.
In this limit
 \eqn\time{t(\tilde \p)={\tilde m_{pl}\over 2\gamma^{1/2}
        T(\tilde \p)^2}\ \ .}
Since $t_c \propto 1/H_c$ and
$t_o \propto 1/H_o$, eqn. \horstar \ is equivalent to the statement that
  \eqn\old{{t(\tilde \p)\over R(\tilde \p)}\gta {t_o\over R_o}\ \ .}

Since $\tilde m_{pl}$ does in fact exceed $M_o$ we see from
eq. \time \ that
the universe is older at a given temperature than in the standard
cosmology.  Writing eq. \time \ in terms of $t_{\rm einst}$, the age of a
cosmology described by Einstein gravity, gives
  \eqn\age{t(\tilde \p)=
        t_{\rm einst}\left({\tilde m_{pl}\over M_o}\right)\ \ }
at a given temperature.
In a standard cosmology, $t_{\rm einst}\sim ({\rm MeV}/T)^2\ {\rm sec}$.
So, at 1 MeV,
$t_{\rm einst}\sim 1$ sec and $t(\tilde \p)\sim 10^7\
{\rm sec}\sim 3 \ {\rm yrs}$.

If, instead, we take $T(\Phi_c\approx \tilde \p)$
to be the temperature of matter/radiation
equality, about 5.5 $\Omega_M h^2$ eV, then
the causality condition requires
        \eqn\ans{\tilde m_{pl}\gta {\cal O}(10^{2})M_o\ \ .}
At $T\sim 1\ $eV,
$t_{\rm einst}\sim 10^{12} {\rm \ sec}\sim 10^5 {\rm \ yrs}$, and
$t(\p)\sim 10^2\  t_{\rm einst}\sim 10^7$ yrs.
We see that the smoothness of the observable universe is resolved
in this
MAD  model by aging the universe.

\centerline{\bf III. D) Problems and Constraints}

The obvious difficulty with this resolution to the
horizon problem is fixing the value of the Planck mass to be $M_o$
by today.
In the
Brans-Dicke model studied here without a potential, the Planck mass
will be hard pressed to make it to the value $M_o$ today.
During the matter dominated era,
$\p$ will initially continue to decrease with time for
$C>0$ and increase with time for $C<0$.$^8$
For $C<0$ then, the Planck mass will only grow larger.
For $C>0$,
it is conceivable that $\p$ will approach
the value of $M_o^2$ during the matter dominated era.  However,
observations constrain the parameter $\om$ to be
$\gta 500$ for a massless Brans-Dicke theory.  The rate at which
$\p$ changes depends on $\om$ and is very suppressed
for large $\om$.  Thus a large $\om$ would confine $\p$ to
near its value at the time of matter radiation equality,
which, as we have seen, may be large.
For example, with $\om=500$ and $\tilde \p^{1/2}=10^2 M_o$
at $T_c \sim 1$ eV, then
today $\p^{1/2}_o\geq 80 M_o$.
This limit can be avoided if there is
a potential for $\p$, e.g. a mass term such as
$V(\psi)={m_{\psi}^2\over 2}\psi^2$.
The interactions measured in the time-delay experiments
fall off rapidly outside the range over which the $\Phi$
field acts.  If $\p$ has an associated mass $m_{\psi}$, then
the range over which $\Phi$ acts $\lambda\sim {1/m_{\psi}}$
could be smaller than the distances over which the
observations are sensitive.  Therefore a massive Brans-Dicke
model would elude observation even if $\om$ is small$^{9}$.
However, in another paper$^{10}$
we discuss the case of Brans-Dicke with a potential
and find that it is difficult to simultaneously
satisfy bounds on time-delay experiments as well
as other constraints.

There are also observations of the rate of change of the gravitational
constant.  These observations impose a much weaker constraint
than the time delay experiments.  They suggest $\om\gta 5$.
For comparison, if $\om=5$ and $\tilde \p^{1/2}=10^2 M_o$, then
one could have
$\p^{1/2}_o\approx M_o$ up to ${\cal O}(1)$.

Another issue of concern is the value
of the Planck mass, and thus the Hubble constant,
during nucleosynthesis.
If $m_{pl}\ne M_o$
during nucleosynthesis, then the
predicted elemental abundances
will be affected.
To resolve the
horizon problem, we found that
the asymptotic value $\tilde m_{pl}=\tilde \p^{1/2}$
had to be much larger than $M_o$.
If, for example, $m_{pl} \simeq \tilde m_{pl} \gg M_o$
during nucleosynthesis,
then $H(\p)\propto {T^2(\p)\over \tilde m_{pl}^{1/2}}$ and the
large Planck mass slows the expansion of the universe.
Consequently, the temperature at which the weak
interactions freese out is lowered, the $n/p$
ratio is maintained at its equilibrium value longer,
and the value of the $n/p$ ratio during nucleosynthesis
is smaller.
This works to decrease the production of ${\rm He}^4$.
Compatibility with observations would then force
$\Omega_b$, the fraction of critical density in
baryonic matter, to be larger.
Actually, since the Hubble
constant can be so much smaller than in the
standard model, a situation may arise where
the weak interactions are still in equilibrium during
the time of ${\rm He}^4$ synthesis; then  the
nucleosynthesis calculations would have to be redone.
We have not investigated the consequences for
abundances of other elements (such as deuterium, lithium,
and He$^3$).  We suspect that, unless the Planck
mass has returned to its present value by the
time of nucleosynthesis,
matching observations on all elements
simultaneously will be impossible.

One could insist that the causality condition is solved
for temperatures greater than an MeV and then invoke a potential to
drive $m_{pl}$ to $M_o$ by the time of nucleosynthesis.
This would also accomodate
the $C=0$ scenario where $\p$ is constant at the value $\tilde m_{pl}$
needed to solve causality. For the
previous results to hold, the potential would have to remain inconspicuous
during the early evolution.  Again, we treat the model
which includes a potential in a separate paper
and find that it is difficult to satisfy all the constraints.

Next we will consider
a more general scalar type theory where $\p(\psi)$ is
such that $\om$ is not constant.  We treat this case below.

\centerline{\bf IV) Case b: $\om\ne$constant:}

\centerline{\bf IV.A) Solution to the Equations of Motion}

Here we extend the analysis to the more general case of $\om$
not constant, again with no potential for the Brans-Dicke field.
We will find the solutions to
$R(\p), H(\p),$ and $T(\p)$
here.  The solutions
during the radiation dominated era
are very similar in spirit to the previous
solutions for $\omega =$ constant.
However, the case of $\omega$ not constant does allow
the possibility of
a small value of $\om$ at early times that matches
onto $\omega \geq 500$ today.
Although we have not yet worked out the evolution
for specific examples of changing $\omega$
during the matter dominated era, one can hope
that $\p$ can change rapidly enough during the period of matter
domination to reach the value $M_o^2$ today.
For certain choices of $\p(\psi)$, $\om(\psi)$ will grow
in order to match the observational
constraints of $\om\gta 500$ today without invoking
a potential. (Actually, the
observational constraints need
to be reinterpreted if $\om\ne$constant.)

If $\Phi$ has any functional form other than
the minimal $\propto \psi^2$, then $\om=8\pi{\p\over(\partial \Phi
/\partial \psi)^2}$
will also be a function of $\psi$.
The equation of motion \one \ in the radiation
dominated era for the case of no potential is
  \eqn\motion{\ddot \p + 3H\dot \p=-{d\om \over d\p}{\dot\p^2\over (3+2\om)}
        =-{\dot \om}{\dot\p\over (3+2\om)}\ \ ,}
so that
  \eqn\mo{\dot \p R^3=-{C\over (1+2\om/3)^{1/2}} \ \ \ \ {\rm and}\ \
        H=-{\ddot \p \over 3\dot\p}-{2\dot \om /3 \over 6(1+2\om /3)}
        \ \ .}
We can use results \motion \ and \mo \ in eqn \two \
as we did in the previous section.  This time,
we define
  \eqn\chiq{\chi = 4 \bar S^{4/3} \gamma C^{-2} \p R^2\ \ }
(this time, there is no factor of $(1 + 2 \omega /3)^{-1}$
in our definition of $\chi$).  Again, we obtain eqn. \fourteen \ .
Since $\omega(\p)$ is as yet unspecified here,
we define the right hand side
of eqn. \fourteen \ to be
  \eqn\sidney{\Sigma_2(\p)=\pm\int_{\p_i}^\p
        {d\p'\over 2\p'}(1+2\om'/3)^{1/2}\ \ .}
Then the solution to \fourteen \ in this case becomes
  \eqn\cici{\chi^{-1/2} = \sinh[\Sigma_2(\p) + {\rm arcsinh} \chi_i^{-1/2}]
        \ \ .}
{}From eqns. \chiq \ and \cici \ , we can find
  \eqn\scale{
        R(\p)={C\over 2\bar S^{2/3}\gamma^{1/2}}{1\over \p^{1/2}
        \sinh[\Sigma_2(\p) + {\rm arcsinh} \chi_i^{-1/2}]}
        \ \ . }
{}From now on we define $\sigma_i = {\rm arcsinh} \chi_i^{-1/2}$
and absorb this term
into the definition of $\Sigma(\p) \equiv \Sigma_2(\p)
+ \sigma_i$.

It follows that
  \eqn\temp{T(\p)={2\bar S\gamma^{1/2}\over C}{\p^{1/2}\sinh\Sigma}
        \ \ .}
For use in the constraint \horstar, we find $H(\p)$ as before,
  \eqn\hphi{H(\p)={4\gamma^{3/2}}\left({ \bar S\over C}\right )^2
        {\sinh^2\Sigma\p^{1/2} \over
        (1+2\om/3)^{1/2}}
        \{ \sinh\Sigma+(1+2\om/3)^{1/2}\cosh\Sigma\}
        \ \ .}
$H(\P)$ can be expressed in terms of the temperature and $\Sigma$
using eq. \temp \ in eq. \hphi.
  \eqn\Htemp{H(\P)=\gamma^{1/2}
        {T^2(\p)\over \p^{1/2}}{1\over (1+2\om/3)^{1/2}}
        \{ \sinh\Sigma+(1+2\om/3)^{1/2}\cosh\Sigma\}
        \ \ .}
The comoving horizon size is
  \eqn\lui{
        {d_{\rm horiz}\over R(\p)}=\left (C\over 4\gamma \bar S^{4/3}\right )
        \left[{1 \over \tanh \Sigma} - {1 \over \tanh \sigma_i}
        \right] \ \ . }

In order to solve the smoothness problem, we need
${1 \over H(\p_c) R(\p_c)} \gta
{1 \over H_o R_o}$ as before.  This gives a constraint
similar to the $\om=$constant scenario.  Written in terms of
the temperature, the constraint is
  \eqn\mass{
        {\p_c^{1/2}\over T(\p_c)}
        {(1+2\om_c/3)^{1/2}\over
        \{ \sinh\Sigma_c+(1+2\om_c/3)^{1/2}\cosh\Sigma_c\}}
        \gta \beta {M_o\over T_o}\ \ , }
where $\beta$ is defined below eq. \hee \
and again the subscript $c$ indicates the values at the time causality
is solved.
Again, the causality condition is
satisfied with the least deviation from Einstein gravity
for small
$\Sigma_c$ (i.e. $\sinh\Sigma_c\sim 0$
and $\cosh\Sigma_c\sim 1$); then eqn. \mass \ becomes
  \eqn\phew{{\p_c^{1/2}\over M_o}\gta \beta {T_c\over T_o}\ \ .}
We see that a large early value of the
Planck mass is again required to solve the causality
condition at high temperatures.

Again, as in the case of pure Brans-Dicke gravity,
$\p$
approaches an asymptotic value $\tilde \p$ from
either above or below.
We can see this by looking at eqn. \mo \ :
for $\omega > 0$,
we can see that $\dot \p \leq C/R^3 \rightarrow
0$ as $R \rightarrow \infty$, and
$\p$ approaches some constant value $\tilde \p$.
For consistency,
we want the scale factor in eqn. \scale \ to grow
very large
as $\p$ approaches $\tilde \p$.
Thus, as $\p \rightarrow \tilde \p$,
we can see from eqn. \scale \ that we need
$\Sigma \rightarrow 0$, i.e.,
$\int_{\p_i}^{\tilde \p} {d \p \over 2 \p}
\left( 1 + 2 \omega /3 \right)^{1/2}
= \mp {\rm arcsinh} (\chi_i^{-1/2})$,
where the upper (lower) sign refers to
$C > (<) 0$.  This
last relation tells us the asymptotic value $\tilde \p$
as a function of initial values $\chi_i$ and $\p_i$.

As an example we take
$\p(\psi)=\hat \p\exp{\left(\psi/\hat \psi\right )}$
where
$\hat \psi$
is a constant mass scale and
$\hat \p$ is a constant with
units of mass squared.
Then
    \eqn\omnot{\om={8\pi{\hat \psi^2\over \p}}\ \ }

The description of this cosmology is very similar to the $\om=$constant
scenario.  Initially, $\dot \p$ is either positive, negative, or zero.
If $\dot \p$ is positive,
then the Planck mass starts small and increases toward
the asymptotic value $\tilde m_{pl}=\tilde \p^{1/2}$.
If $\dot \p$ is negative,
then the Planck mass begins large and
decreases toward
$\tilde m_{pl}$.
To satisfy the causality condition
when the Planck mass is at the asymptotic value $\tilde m_{pl}$,
we need $\tilde m_{pl}/M_o\gta\beta
\tilde T/T_o$.

Although this yields a very similar description of
the early universe as the $\om$ constant scenario
did, there is one advantage.  Since
$\om\propto \p^{-1}$,
$\om$ is small for large values of $\p$. For small $\om$,
$\p$ changes rapidly with time when the universe is
dominated by nonrelativistic matter.
This buys $\p$
the chance to move toward $M_o^2$
by today.  As $\Phi$ decreases, $\om$ increases
and automatically turns off the change in $\p$.
A more thorough treatment of the equations of motion during matter
domination with $\om\ne$constant is needed to verify this
suggestion.

\centerline{\bf V) Conclusions}

During a MAD  era,
the Planck mass is large and the universe
is older at a given temperature than in a standard cosmology.
The comoving size of a causally connected region
$1/H R$
is correspondingly
larger.
Since the observed universe only reaches out to about
recombination (or possibly back to nucleosynthesis),
a rough estimate of the comoving radius of
the observable universe is the present comoving
Hubble radius $H^{-1}_o/R_o\sim \Delta t_o/R_o$,
where $\Delta t_o$ is roughly the time elapsed since the universe became
matter dominated.
The smoothness of the observable universe
can be explained if the universe ages sufficiently so that
the comoving horizon size at the time of
nucleosynthesis  is
$H^{-1}/R\gta \Delta t_o/R_o$.
Observations of the age of the universe
from the Hubble diagram, nucleocosmochronology, and
ages of globular clusters only place limits on the age
of the universe subsequent to the time when stars formed;
thus, the universe may in fact become much older in
the radiation dominated era than one would expect
from the standard model.

We found that a MAD era with
large Planck mass ages the universe
sufficiently to resolve the horizon problem. For a Brans-Dicke
theory without a potential
during the radiation dominated era, the Planck mass approaches an
asymptotic value, $\tilde m_{pl}$.
This asymptotic value can be chosen to
satisfy the causality condition at a specified temperature $T_c$:
$\tilde m_{pl}/M_o\gta T_c/T_o$.
For instance, for causality to be satisfied by the time of
nucleosynthesis, we need the asymptotic value $\tilde m_{pl}
\gta 10^7M_o$.  Or, for causality to be satisfied by the time
of matter/radiation equality, we need $\tilde m_{pl}\gta 10^2 M_o$.
If the Brans-Dicke parameter
$\om$ is greater than 500 and constant, $m_{pl}$
will be unable to reach the value $M_o = 10^{19}$ GeV
by today.  Also,
if the Planck mass is large during primordial
nucleosynthesis, then
the standard big bang nucleosynthesis predictions will be altered.
We suggest that $m_{pl}$ may be driven down to almost its
present value by the time of nucleosynthesis
if there is a potential in the theory for $\p$
(however, in a separate paper$^{10}$ we show that
such an approach is typically overconstrained)
or if $\om$ is allowed to vary as a function of time.
(In a separate paper we also discuss the flatness
problem and the monopole
problem.)$^{10}$

For a more general scalar theory of gravity than the Brans-Dicke proposal,
where $\om$ is not constant, we find again that the Planck mass can
approach an asymptotic value as the universe evolves.
As before, if
$\tilde m_{pl}/M_o\gta T_c/T_o$,
then the smoothness of the present
observable Hubble volume can be explained in this model.
In addition, if $\om\ne$constant, then the Planck
mass may change rapidly enough with time to reach the value of $M_o$ by
today. The equations of motion during the matter dominated era for a more
general scalar theory need to be
studied in detail, as do the constraints on general models.

\bigskip
\centerline{\bf Acknowledgements}
\medskip

We would like to thank Fred Adams, Alan Guth,
and Helmut Zaglauer for
helpful conversations.
KF  and JJL thank each other for getting MAD.
KF thanks
the ITP at U.C. Santa Barbara and the
Aspen Center for Physics, where part of
this work was accomplished,
for hospitality.
We acknowledge support from NSF Grant No. NSF-PHY-92-96020, a Sloan
Foundation fellowship, and a Presidential Young Investigator award.

%
%

\centerline{\bf Appendix}

We present here the solutions to the equations of motion
during the radiation dominated era for
a Brans-Dicke theory with $\kappa=\pm 1$.  The $\p$ equation of motion
reduces to $\ddot \p+3H\dot \p=0$ so that, as before,
  \eqn\again{\dot \p R^3=-C \ \ \ {\rm and}\ \ \
H=-{\ddot \p\over 3\dot \p}\ \ \  .}
Solving equation
\two\ for $H$ with $C\ne0$ and $\kappa\ne 0$ gives
  \eqn\withk{H=-{\dot \p\over 2\p}\left [1+ 2\ep
        \sqrt{1+\chi-Q^2\chi^2}\right ]\ \ }
where $\ep=\pm{(1+2\om/3)^{1/2}\over 2}$, $\chi$ is defined as in
\twelve \
and
  \eqn\q{Q^2={\ep^2 C^2\over \gamma^2\bar S^{8/3}}\kappa\ \ .}
Using $H=\dot R/R$, the definition of $\chi$, and rearranging,
we are left
with the integral
  \eqn\appendint{\int_{\chi_i}^\chi
{d\chi'\over \chi'\sqrt{1+\chi'-Q^2\chi'^2}}=\mp
        \int_{\Phi_i}^\Phi \left (1+{2\omega\over 3}\right)^{1/2}
        {d\Phi' \over \Phi'}\ \ .}
Integrating this equation and using
$R=(\ep C/\bar S^{2/3})(\gamma\p)^{-1/2}\chi^{1/2}$, from the
definition
of $\chi$, we find,
   \eqn\curv{R={\ep C\over \bar S^{2/3}\gamma^{1/2}}
     {1\over \p^{1/2}}
        \left\{ {1 \over
     \sinh^2\Theta  +Q^2\exp{(-2\Theta)}}
\right \}^{1/2}\ \ ,}
where, as before, $\Theta  =\ep\ln(\p/\tilde \p)$.
The temperature of the universe is found from adiabaticity to be
        \eqn\toe{T={\bar S\gamma^{1/2}\over \ep C}\p^{1/2}
        \left\{ {
     \sinh^2\Theta  +Q^2\exp{(-2\Theta)}}
\right \}^{1/2}\ \ .}
The causality condition becomes
\eqn\causing{{\p_c^{1/2}\over T_c}2\ep\left [{
(\sinh^2\Theta_c+Q^2\exp({-2\Theta_c}))^{1/2}\over
\sinh^2\Theta_c+2\ep\sinh\Theta_c\cosh\Theta_c+Q^2(1-2\ep)
\exp({-2\Theta_c})
}\right ]\gta \beta {M_o\over T_o}\ \ .}
Notice that as $Q^2\rightarrow 0$ \curv, \toe,
and \causing\ reduce to the corresponding results
for a flat
universe.  Similarly, for large $\Theta$,
$e^{-2\Theta}\rightarrow 0$, and
we have the
same causality condition as in the case of the
flat universe.
We will further discuss the case of nonzero curvature,
together with the issue of flatness, in another paper.

\vskip 1.0truein
\centerline{\bf REFERENCES}
\vskip 0.10truein

\item{[1]} A. H. Guth, {\it Phys. Rev.} D {\bf 23}, 347 (1981).

\item{[2]} C. Brans and C. H. Dicke, {\it Phys. Rev.}
{\bf 24}, 925 (1961); S. Weinberg, {\it Gravitation and Cosmology:
Principles and Applications of the General Theory of Relativity},
(New York: John Wiley and Sons) (1972).

\item{[3]} See the discussion in C. M. Will,
{\it Theory and Experiment in Gravitational Physics},
(New York: Cambridge University Press) (1981).

\item{[4]} D. La and P. J. Steinhardt, {\it Phys. Rev. Lett.}
{\bf 376}, 62 (1989); D. La and P. J. Steinhardt,
{\it Phys. Lett.}, {\bf 220 B}, 375 (1989).

\item{[5]} P. J. Steinhardt and F. S. Accetta,
{\it Phys. Rev. Lett.} {\bf 64}, 2740 (1990).

\item{[6]} P.G. Bergmann, {\sl Int. J. Theor. Phys.}
{\bf 1}, 25 (1968).

\item{[7]} R.V. Wagoner, {\sl Phys. Rev. D} {\bf 1},
3209 (1970).

\item{[8]} S. Weinberg, {\it Gravitation and Cosmology:
Principles and Applications of the General Theory of Relativity}
(New York: John Wiley and Sons) (1972).

\item{[9]} Helmut Zaglauer, private communication.

\item{[10]} K. Freese and J.J.Levin, preprint.

\bye